\documentclass[a4paper,superscriptaddress,aps,pre,nolongbibliography,twocolumn,showpacs,floatfix]{revtex4-1} 

\usepackage{graphicx,helvet}
\usepackage{color}
\usepackage{bm}                      
\usepackage{tikz}
\usepackage{amsfonts}
\usepackage{lipsum}
\definecolor{mygray}{gray}{0.4}
\definecolor{light-blue}{rgb}{0.8,0.85,1}
\usepackage{amsmath}%
\usepackage{MnSymbol}%
\usepackage{wasysym}%
 \usepackage{soul}

\usetikzlibrary{decorations.shapes}

\newcommand{\beqa}{\begin{eqnarray}}
\newcommand{\eeqa}{\end{eqnarray}}
\newcommand{\beq}{\begin{equation}}
\newcommand{\eeq}{\end{equation}}

\usepackage{amsmath}

\def\idmat{\mathbb{I}}

\newcommand{\tr}{\mathop{\mathrm{Tr}}\nolimits}

\newcommand{\ket}[1]{{\vert #1 \rangle}}
\newcommand{\bra}[1]{{\langle #1 \vert}}
\newcommand{\braket}[2]{{\langle #1\vert #2 \rangle}}

\newcommand{\ifimar}{Instituto de Investigaciones F\'isicas de Mar del Plata (IFIMAR), CONICET-UNMdP,  Mar del Plata,  Argentina}
\newcommand{\conicet}{Consejo Nacional de Investigaciones Cient\'ificas y Tecnol\'ogicas (CONICET), Argentina}
\newcommand{\uba}{Departamento de F\'isica ``J. J. Giambiagi'' and IFIBA, FCEyN, Universidad de Buenos Aires, 1428 Buenos Aires, Argentina}
\begin{document}
\title{Semiclassical approach to the work distribution}
\author{Ignacio Garc\'ia-Mata} \affiliation{\ifimar} \affiliation{\conicet}
\author{Augusto J. Roncaglia}\affiliation{\uba}
\author{Diego A. Wisniacki} \affiliation{\uba}

\begin{abstract}
Work in closed quantum systems is usually defined by a two-point  measurement. 
This definition of work is compatible with quantum fluctuation theorems but it fundamentally differs from its classical counterpart.
In this paper, we study the correspondence principle in quantum chaotic systems. 
We derive a semiclassical expression of the work distribution for chaotic systems undergoing a general, finite time, process.  
This semiclassical distribution converges to the classical distribution in the usual classical limit. We  
show numerically that, for a particle inside a chaotic cavity, the semiclassical distribution provides a good approximation to quantum distribution.
 \end{abstract}
%
%
\maketitle
\section{Introduction}
The last years have seen growing interest in studying the thermodynamics of small quantum systems away from equilibrium \cite{PolkovnikovRMP2011,Goold16}. 
Thus, several thermodynamical quantities defined in the classical framework were extended to the quantum regime.
In particular, the definition of quantum work is rather subtle \cite{Talkner07,Campisi2011,talkner2016aspects}. 
For driven (but otherwise isolated) systems, work is related to the change of energy. 
Thus quantum work is generally defined by two-point energy measurements, 
one at the beginning and the other at the end of the process \cite{Tasaki00,Kurchan2000,Talkner07}.
Quantum work is then a stochastic quantity obtained as the difference between the two energy measurement outcomes.
Notably, the quantum version of non-equilibrium work relations 
\cite{Jarzynski1997,Tasaki1998,Crooks1999}, that have been shown to be very useful in the estimation of free energy difference for small classical systems
\cite{Hummer2001,Collin2005,Liphardt2002}, follow directly from this definition of quantum work \cite{Tasaki00,Kurchan2000,mukamel2003quantum}.
However, since the quantum work relies on projective measurements without a classical counterpart, 
it could be argued that its definition is tailored so that it satisfies the above relations.   

In order to investigate the relationship between the quantum and classical definitions of work, it has been 
suggested to study the correspondence principle between those distributions \cite{Jarzynski2015}. 
This correspondence between quantum and classical work distributions was shown to exist
for integrable systems (1D) \cite{Jarzynski2015,Deffner10}, a chaotic system \cite{Quan2016}
and a many-body system \cite{Wang2017}.  In a recent work \cite{NAD2017}, this question was also addressed for general chaotic systems after
an instantaneous process (quench). There, a semiclassical expression for the characteristic function of the work distribution is introduced.
More importantly, it is shown that the resulting semiclassical work distribution not only provides a good approximation 
to the quantum distribution at high temperatures for 
a particle inside a billiard, but it is shown analytically that in the usual classical limit $\hbar\to 0$ it approaches the classical distribution.
The derivation is based on three key ingredients that are obeyed by generic chaotic systems: 
the connection between the characteristic function with the Loschmidt echo \cite{Silva2008, DiegoScholar}, the semiclassical dephasing representation for the fidelity amplitude  \cite{vanicek2003,vanicek2004,vanicek2006}  and the Berry-Voros quantum ergodic conjecture \cite{Berry_1977,Voros}, which states that the Wigner functions of eigenstates of chaotic systems are peaked on the corresponding energy shell. 
In this paper, we show that a semiclassical expression can also be found for a more general process that occurs in finite time. This distribution
has  the correct classical limit and also provides a good approximation to the quantum distribution, supporting the definition of quantum work via the two-point measurement scheme.

\section{Semiclassical approach}
Let us consider a system which evolves under a time dependent Hamiltonian $H_t$.
Initially, the system is a thermal Gibbs state  \mbox{$\rho_\beta=\exp(-\beta H_0/Z_0^{\rm Q})$} at temperature $\beta^{-1}$,
where $Z_0^{\rm Q}=\tr[\exp(-\beta H_0)]$ is the partition function.  Suppose a projective energy measurement is performed and yields $E_0^m$ as a result.
After that, the system is subjected to a process characterised by a time-dependent Hamiltonian $H_t$, 
where a parameter of the Hamiltonian is switched at a finite rate during a time $\tau$, leading $H_0$ to $H_\tau$.  
This process induces an evolution described by a unitary transformation $U_\tau\equiv \mathcal{T} e^{-\frac{i}{\hbar} \int_0^\tau \, H_t\,dt}$. At time $\tau$ a second energy measurement is carried out, giving $E_\tau^n$ as a result. 
The quantum work after this process is related to the change of energy as $W\equiv E^n_\tau-E^m_0$.
The procedure we described is known as the two-point measurement scheme, and defines 
quantum work as a stochastic quantity characterised by a probability distribution \cite{Kurchan2000,Tasaki00,Talkner07}:
\begin{equation}
P^{\rm Q}(W)=\sum_{n,m}P^{\rm Q}(m) P^{\rm Q}(n| m)\delta[W-(E^n_\tau-E^m_0)],
\label{PWQ}
\end{equation}
where ${P^{\rm Q}(m)={e^{-\beta E^m_0}}/{Z^{\rm Q}_0}}$ is the probability that the initial energy measurement
yields $E^m_{0}$ as a result,  ${P^Q(n|m)=|\bra{\phi_\tau^n} U_\tau \ket{\phi_0^m}|^2}$ is the conditional probability to obtain 
$E^n_{\tau}$ at the final measurement given that the initial result was $E^m_{0}$, and $\delta$ 
is the Dirac $\delta$-function ($\ket{\phi_t^m}$ are the eigenstates of energy $E^m_t$ of $H_t$). In this case,  we assume that both Hamiltonians have non-degenerate spectrum.
Several methods to measure experimentally this distribution have been proposed  
\cite{Huber2008,Dorner2013,Mazzola2013,campisi2013employing,Roncaglia2014}, and some of them have been recently implemented 
verifying fluctuations theorems in the quantum regime  \cite{Batalhao2014,An2014,cerisola2017using}. 

The semiclassical approach that we propose is based on the study of the characteristic function,
which is defined as the Fourier transform of the work probability distribution:
\begin{equation}
G(u)=\int dW e^{i u W} P(W).
\end{equation}
Remarkably, in the quantum case, the characteristic function  can also be expressed as a correlation function\cite{Talkner07}:
\beq
G^{\rm Q}(u)=\tr [e^{i u H^{\rm H}_{\tau}} e^{-i u H_{0}}\rho_{\beta}],
\label{eq:GQ}
\eeq
where $H^{\rm H}_\tau=U^\dagger_\tau H_{\tau} U_\tau$ is the Hamiltonian $H_\tau$ in the Heisenberg picture.
In this way, $G^{\rm Q}(u)$ can also be viewed as the amplitude of a Loschmidt echo \cite{Silva2008} (or fidelity amplitude),
where the forward evolution is governed by the Hamiltonian $H_{0}$ and the backward evolution by $H^{\rm H}_{\tau}$.

 A semiclassical expression of the characteristic function for instantaneous processes ($U_\tau = \idmat$)  was recently \cite{NAD2017} proposed.
The approach in \cite{NAD2017} is based on the dephasing representation (DR) \cite{vanicek2003,vanicek2004,vanicek2006}, which
provides a semiclassical expression to the fidelity amplitude.  
The DR avoids fundamental problems of traditional semiclassical approaches \cite{vanicek2004} and it has also the advantage that in most cases it is the most 
 efficient to compute numerically  \cite{Zambrano2013}.
A key assumption in the semiclassical derivation is that the echo evolution,  given by Eq.~\eqref{eq:GQ}, is governed 
by two Hamiltonians with a clear classical counterpart. While we assume this true for $H_{0}$ and $H_{\tau}$, 
one can notice that for $H^{\rm H}_\tau$ and an arbitrary process $U_\tau$ in general this will not be the case.
Therefore, bellow we  propose a classical Hamiltonian for $H^{\rm H}_\tau$. First, let us notice 
that  $H^{\rm H}_\tau$ and $H_{\tau}$ have the same spectrum since they are related by a unitary transformation, 
and also the eigenstates of $H^{\rm H}_\tau$ are the eigenstates of $H_{\tau}$ evolved 
with the backward process $U^\dagger_\tau$.  Based on this observation, we consider the following classical Hamiltonian associated with $H^{\rm H}_\tau$ for the DR: 
$H^{\rm C}_{\tau}(z)\equiv H_{\tau}(z_\tau(z))$, where $z$ denotes a  phase space  point ($z\equiv(q,p)\in\mathbb{R}^{2D}$,  $D$ is the number of degrees of freedom) 
and $z_\tau(z)$ denotes the final phase space point of a trajectory that evolves during time $\tau$ from an initial condition $z$ with the Hamilton's equations during
the process. Our ansatz is also supported by the fact that this Hamiltonian association is correct in some cases, for instance quadratic 
Hamiltonians with  $U_\tau$ described by symplectic transformations, and also the resulted expression for the probability of work has the correct classical limit. 
We will also show that the DR, using this classical Hamiltonian, provides a good approximation to the quantum distribution of wok.
\begin{figure}
\centerline{\includegraphics[width=0.55\linewidth]{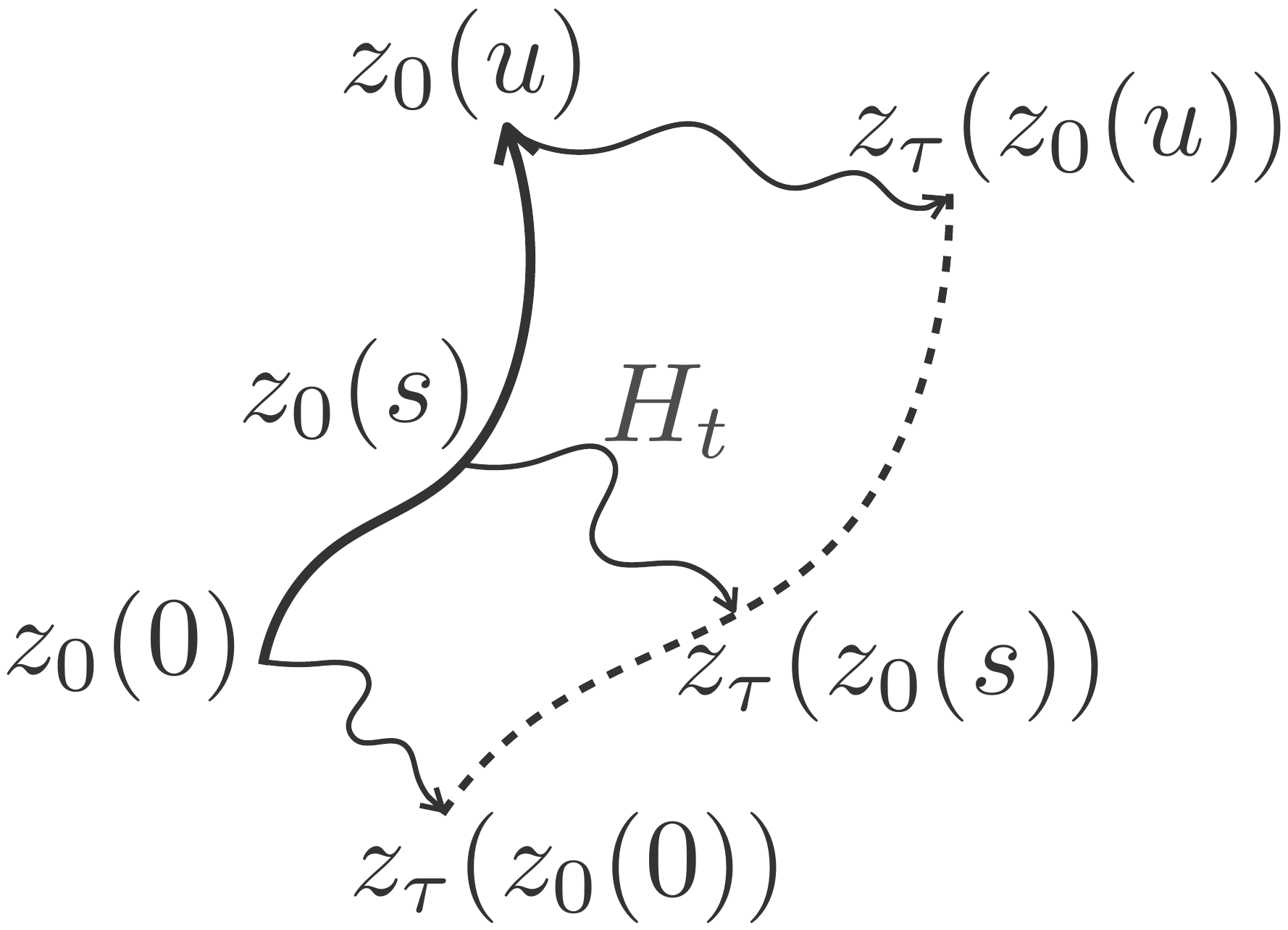}}
\caption{Schematic depiction of the phase trajectories for the calculation of the points $z_\tau(z_0(s))$ to be used in the calculation of 
$G^{\rm SC}(u)$. The solid trajectory $z_0(s)$ is obtained via the classical evolution generated by the Hamiltonian $H_0$ with initial condition $z_0$.
The dashed trajectory $z_\tau(z_0(s))$ is obtained by evolving each phase space point $z_0(s)$ with the classical process during time $\tau$.
For the calculation of $G^{\rm SC}(u)$, the Hamiltonian $H_0$ is integrated along the solid trajectory 
while $H_\tau$ is integrated along the dashed one.  
\label{fig:schematic}}
\end{figure}
Thus, using these ideas and following the same procedure as in Ref.~\cite{NAD2017}, we arrive at the semiclassical expression 
for the characteristic function:
\beq
\label{GSC}
G^{\rm SC} (u)=\int d^D z_0 \; {\cal W}_\beta(z_0)\; e^{\frac{i}{\hbar} \,\Delta S_\tau(z_0,u \hbar)}
\eeq
where ${\cal W}_\beta(z_0)$ is the Wigner function of the thermal state $\rho_\beta$.
The fundamental difference with the quench process \cite{NAD2017} resides in the action difference
\beq
\label{DeltaS}
\Delta S_\tau(z_0, u \hbar) \equiv \int_0^{u\hbar} [H_{\tau}(z_\tau(z_0(s)))-H_{0}(z_0(s))]\,ds,
\eeq
where $z_0(s)$ denotes the phase space coordinate at time $s$ of a trajectory generated by a classical Hamiltonian 
$H_{0}$ with initial condition $z_0$. 
While for a quench both Hamiltonians that appear in the action difference are evaluated at the same point $z_0(s)$, for a general process
the Hamiltonian at time $\tau$ is evaluated at $z_\tau(z_0(s))$. In Fig.~\ref{fig:schematic} we show a schematic representation of this transformation.

The final expression for $G^{\rm SC}$ is obtained after expressing the Wigner
function for the thermal state. 
Thus, we consider that $H_{0}$ is chaotic and we use the quantum ergodic conjecture (QEC) \cite{Berry_1977, Voros}. Following the same
procedure as in Ref.~\cite{NAD2017}, we arrive at a semiclassical expression of the characteristic function:
 \beq
\label{GSCfinal}
G^{\rm SC}(u)=\int {d^{2D} z_0}\, \frac{e^{-\beta H_{0}(z_0)}}{Z^{\rm C}_{0}}  e^{\frac{i}{\hbar} \,\Delta S_\tau(z_0,u \hbar)},
\eeq
where $Z^{\rm C}_{0}~=~\int dz\, \exp[-\beta \,H_{0}(z)]$ is the classical partition function.

\section{Classical limit}
We have obtained  a semiclassical expression for the characteristic function, from which we can derive the semiclassical approximation of the work probability distribution. 
One of the key features of $G^{\rm SC}(u)$ is its parametrical dependence on the effective Planck constant 
(the classical limit being $\hbar\to 0$).  

Now we show that using this semiclassical approximation one can obtain the classical work distribution in the usual semiclassical limit
$\hbar\rightarrow 0$. It is easy to check that in this limit the exponential of Eq. \eqref{GSCfinal} tends to
\beq
G_{\hbar\rightarrow 0}^{\rm SC}(u) =  \int {d^{2D}z_0} \, \frac{e^{-\beta H_{0}(z_0)}}{Z^{\rm C}_{0}} e^{i[H_{\tau}(z_\tau(z_0))- H_{0}(z_0)]u}.
\eeq
We can now include two integrals in energies, as it is done in Ref.~\cite{NAD2017},
\begin{widetext}
\beq
G^{\rm SC}_{\hbar\rightarrow 0}(u) =
 \int {d^{2D}z_0} \int dE_{\tau}\, dE_{0} \;
 \frac{e^{-\beta E_{0}}}{Z^{\rm C}_{0}} \; \delta[E_{\tau} - H_{\tau}(z_\tau(z_0))] \;
  \delta[E_{0} -  H_{0}(z_0)] \, e^{i(E_{\tau} -E_{0})u}.
\eeq
In order to obtain the probability of work we perform the Fourier transform. Finally after
multiplying and dividing by $g_{0}(E)$ (notice that $g_{0}(E)={\int d^{2D}z\; \delta(E-H_{0}(z))}$ is the density of states), and interchanging the order of the integrals,
we arrive at:
\beqa
P^{\rm SC}_{\hbar\rightarrow 0}(W) &=&  \int dE_{0} \; \frac{e^{-\beta E_{0}}g(E_{0})}{Z^{\rm C}_{0}}
\int dE_{\tau} \int {d^{2D}z_0}\; \frac{\delta[E_\tau - H_{\tau}(z_\tau(z_0))] \, \delta[E_{0} -  H_{0}(z_0)]}{g_{0}(E_{0})} \, \delta[W-(E_{\tau} -E_{0})] \nonumber \\
&=& \int dE_{0} \; \bar P_{0}^{\rm C}(E_{0}) \int dE_{\tau} \; \bar P^{\rm C}(E_{\tau} |E_{0})\, \delta[W-(E_{\tau} -E_{0})] \nonumber \\
&=& P^{\rm C}(W)
\eeqa
\end{widetext}
where $\bar P^{\rm C}_{0}(E)=e^{-\beta E}g_{0}(E)/{Z_{0}^{\rm C}}$, 
$\bar P^{\rm C} (E_\tau|E_0) =\int dz_0\,{\delta[E_\tau - H_{\tau}(z_\tau(z_0))]\,\delta[E_0-H_{0}(z_0)] }/g_{0}(E_0)$, and $P^{\rm C}(W)$ is the
classical probability distribution of work (see Ref.~\cite{Jarzynski2015}).

\begin{figure}[htb]
\centerline{\includegraphics[width=0.95\linewidth]{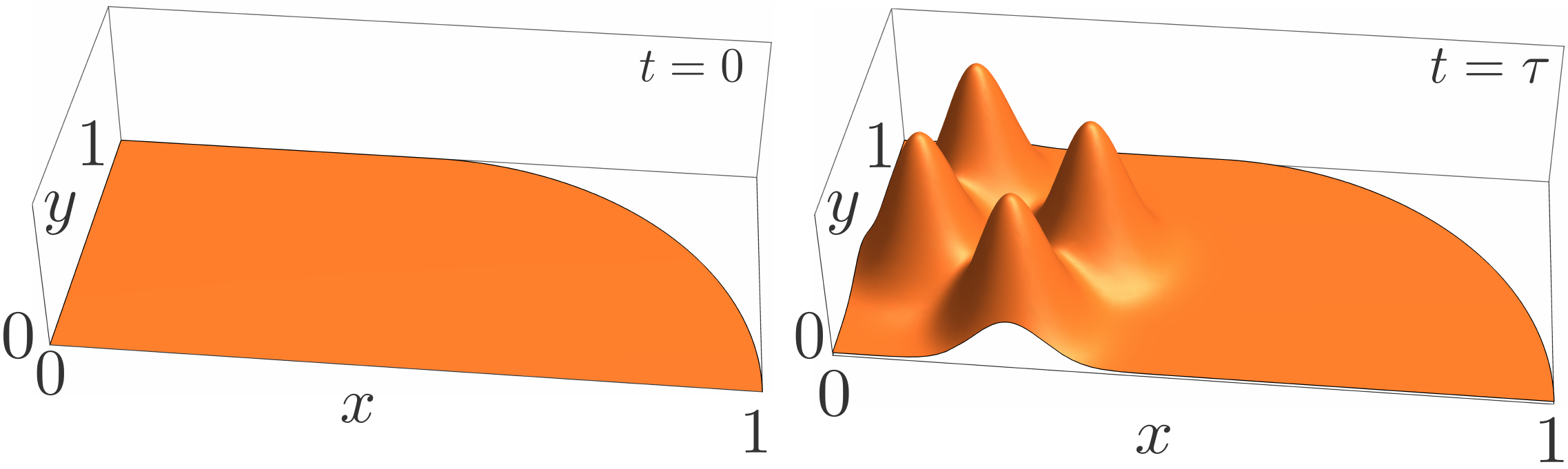}} 
\caption{Representation of the potential that inceases during the process from $t=0$ to $t=\tau$. 
\label{fig:gaussianas}}
\end{figure}

\begin{figure*} 
\centerline{\includegraphics[width=0.95\linewidth]{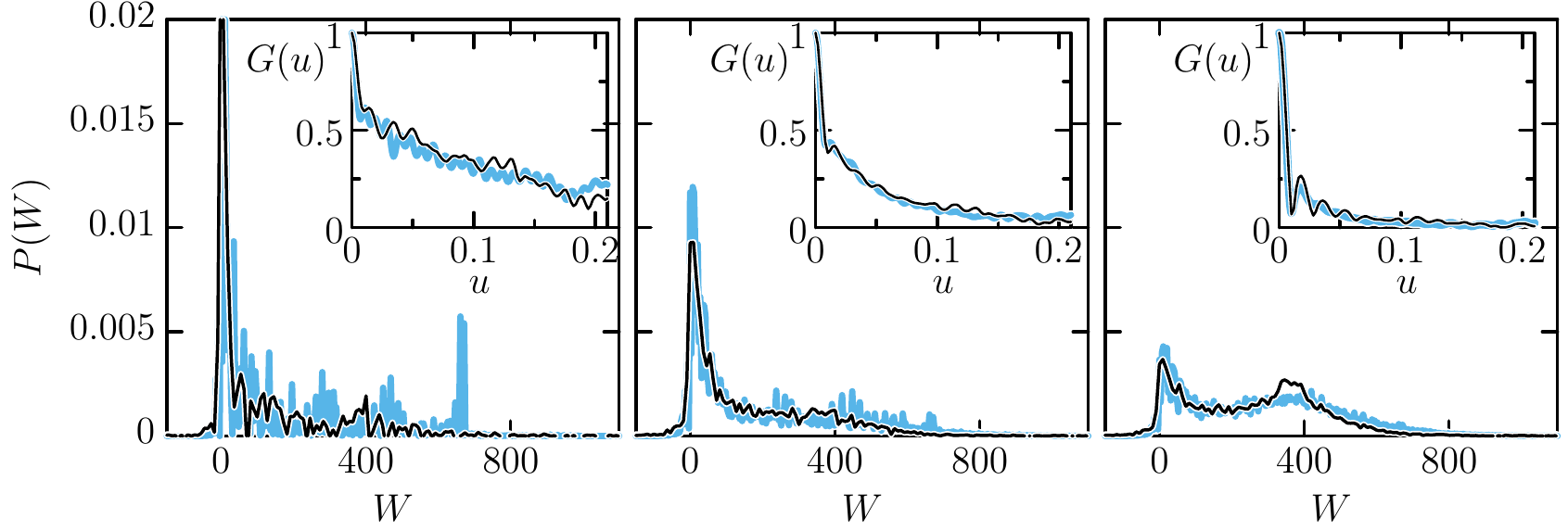}}
\caption{Main panels show the work distribution for $\lambda=180$,  $\sigma=0.1$ and $\tau=0.1$ with temperatures: left $\beta^{-1}=2^5$, center $\beta^{-1}=2^8$, and right $\beta^{-1}=2^{10}$. Solid black lines correspond to the semiclassical calculation and
 the light-blue lines correspond to the quantum calculation.
The insets show the characteristic function obtained using Eq.~\eqref{GSCfinal}. 
\label{fig:pwgu}
}
\end{figure*}

\section{Numerical example}
Now we show a specific example where this semiclassical  expression of the characteristic function is evaluated and compared with its quantum counterpart.
In particular we consider a paradigmatic example of quantum chaotic studies:  a quantum particle inside a stadium billiard (hard walls and desymmetrized), with 
mass $m=1/2$, radius $r=1$, and straight line of length $l=1$. Thus, the thermodynamic process is characterised by the following time-dependent Hamiltonian 
\begin{equation}
H_{t}=H_{0}+\frac{t}{\tau}\, V(x,y)
\label{eq:Ht}
\end{equation}
where $H_0$ is the stadium billiard Hamiltonian, $x$ and $y$ are position coordinates in a bidimensional space ($q\equiv(x,y)$),
and $V(x,y)$ is a smooth potential consisting on the sum of four Gaussians
\begin{equation}
V(x,y)= \lambda \sum_{i=1}^4\frac{1}{2\pi\sigma^2}  e^{-\frac{(x-x_i)^2+(y-y_i)^2}{2 \sigma^2}},
\end{equation}
of width $\sigma$, and strength $\lambda$ (see Fig.~\ref{fig:gaussianas}).
The Gaussians are centered at
$(x_1,y_1)=(0.2,0.4)$, $(x_2,y_2)=(0.67,0.5)$, $(x_3,y_3)=(0.5,0.15)$ and  $(x_4,y_4)=(0.3,0.75)$. 
During the evolution, the perturbation $V(x,y)$ increases linearly in time from $t=0$ to $t=\tau$.
Thus, the inverse of the parameter $\tau$ represents the speed of the process. 
\begin{figure}
\centerline{\includegraphics[width=0.95\linewidth]{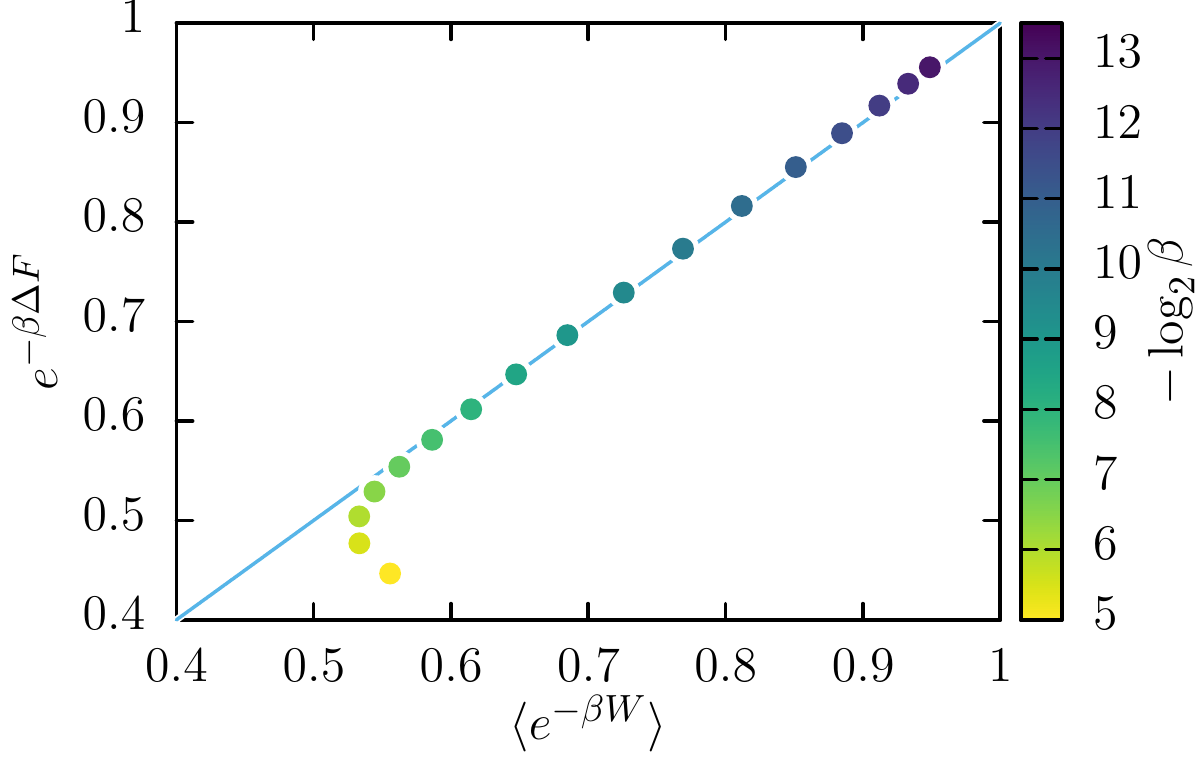}}
\caption{Jarzynsky relation for the process described in the text for different temperatures.  The width of the Gaussians that appear in the potential  $\sigma=0.1$, their final heigh $\lambda=180$, and the duration of the process $\tau=0.1$. See text for more details.
\label{fig:jarz}}
\end{figure}

In order to compute $G^{\rm SC}(u)$, determined by Eq.~(\ref{GSCfinal}), first
we choose $z_0$ as a random initial position and momenta inside the unperturbed stadium. Then, we obtain the unperturbed trajectory $z_0(s)$
using an efficient geometrical algorithm. Finally, each $z_0(s)$ is used as initial condition for the classical evolution generated by 
$H_{t}$, from $t=0$ to $t=\tau$, obtaining $z_\tau(z_0(s))$ (see Fig.~\ref{fig:schematic}).  
The classical time-dependent evolution is obtained using a fourth order Runge-Kutta method. 
After computing $G^{\rm SC}(u)$, the semiclassical work distribution is then directly obtained by evaluating the Fourier transform. 

We compare the semiclassical calculation with
the quantum probability distribution $P^{\rm Q}(W)$ of Eq.~\eqref{PWQ}.
This requires the evaluation of ${P^{\rm Q}(n|m)=|\bra{\phi_\tau^n} U_\tau \ket{\phi_0^m}|^2}$ and we proceed as follows. 
The eigenstates of $H_0$ are obtained using the scaling method \cite{vergini95}. The eigenstates $\ket{\phi_\tau^n}$ are obtained 
by  diagonalisation of $H_\tau$ in the unperturbed basis.
Due to the time consuming limitations of the numerical integration, in our numerics the maximum energy level was fixed at $\approx 1400$, corresponding  wavenumber $k_{\rm max} \approx 100$. This also determined the maximum temperature that we consider in our simulations.
For the calculation of $U_\tau\ket{\phi_0^m}$ we write the time-dependent Schr\"odinger equation for a generic state 
$\ket{\psi_t}=\sum_l a_l(t)\,\ket{\phi_0^l}$:
\begin{eqnarray}
i\partial_t\ket{\psi_t}&=&H_t\ket{\psi_t}\ \ \ \ (\hbar=1)\\
i\sum_l \dot{a}_l(t)\ket{\phi_0^l}&=&\sum_l a_l(t)\left(E_0^l+ \frac{t}{\tau} V(x,y)\right)\ket{\phi_0^l}.\nonumber
\end{eqnarray}
Then we obtain a system of coupled equations
\begin{equation}
\dot{a}_k(t)=-i \left[a_k(t) E_0^k +\frac{t}{\tau}\sum_l a_l(t)\,\bra{\phi_0^k}V(x,y)\ket{\phi_0^l}\right],
\label{Difeq}
\end{equation}
that we solved numerically, with initial conditions
 $a_m(0)=1$ y $a_{l\ne m}(0)=0$ in  (\ref{Difeq}). After that,  
$U_\tau\ket{\phi_0^m}\equiv\ket{\psi_\tau}=\sum_l a_l(\tau)\,\ket{\phi_0^l}$, so 
\begin{equation}
P^{\rm Q}(n|m)=|\bra{\phi_\tau^n}U_\tau \ket{\phi_0^m}|^2=|\sum_l \braket{\phi_\tau^n}{\phi_0^l}\,a_l(\tau)|^2.
\end{equation}

In Fig.~\ref{fig:pwgu} (main panels) 
we show the probability distribution of work using the semiclassical calculation along with the quantum one. 
For $P^{\rm SC}(W)$ we first compute $G^{\rm SC}(u)$ and then the Fourier transform is performed. 
There we show results for three three different values of $\beta^{-1}=2^5,\,2^8,\,2^{10}$.  The 
semiclassical calculation provides a better approximation to the quantum result as the temperature increases (small values of $\beta$). 
For large values of $\beta$, only a few low lying energy eigenstates contribute to the probability distribution, and  it is expected  that the Berry-Voros conjecture 
does not hold.  For instance, at $\beta^{-1}=2^{5}$, due to the fact that the mean level spacing for the unperturbed billiard is $\Delta E\approx 7$, the number of 
energy levels that are contributing is rather small ($\approx 10$). 
We should also point out that our semiclassical calculation gives better results for the quench  (see Ref.~\cite{NAD2017})  
than for the continuous  process we describe here.  
This is mainly due to the fact that, in this case, the semiclassical approach relays in the approximation 
$H^{\rm C}_{\tau}(z)\equiv H_{\tau}(z_\tau(z))$ and this may not be accurate for arbitrary transformations. 
Our simulations suggest that as we decrease the parameter $\tau$ (a measure of adiabaticity) the semiclassical approximation becomes better.

We benchmark the accuracy of our expression by evaluating the Jarzynzki identity \cite{Jarzynski1997}. 
If the system is initially at thermal equilibrium, the Jarzynski identity reads: 
\beq
\label{jarz}
\langle e^{-\beta W}\rangle=e^{-\beta\Delta F},
\eeq
where $\Delta F=-\ln (Z^{\rm Q}_\tau/Z^{\rm Q}_{0})/\beta$ is the change in the Helmholtz free energy, and the angular brackets denote
an average with respect to the work probability distribution of Eq.~\eqref{PWQ}. 
This relation states that for a thermal initial state  the mean value of a function of work (a non-equilibrium 
quantity) is determined by the free energy difference (an equilibrium quantity). 
Thus, transformations that are arbitrary away from equilibrium contain information about equilibrium quantities. 
In Fig.~\ref{fig:jarz} we show Eq.~\eqref{jarz} evaluated numerically. The left side of 
Eq. \eqref{jarz} is calculated from the semiclassical probability of work.
While the right side of Eq.~\eqref{GSCfinal} is obtained by calculating  the quantum energies of $H_0$ and $H_{\tau}$ 
for different values of $\beta$. The main source of errors in this case came from the estimation of the weighted assigned to negative
values of work \cite{jarzynski2006rare,gore2003bias}.
Here again it is clear that in the region where the approximations are valid, i.e. small $\beta$, the Jarzynski identity is satisfied with a reasonable
error.  

\noindent
\section{Summary}
The definition of quantum work via the two-point measurement scheme could 
be challenged as being arbitrarily defined in order to satisfy the non-equilibrium fluctuation relations. 
In order to settle this controversy it is expected that correspondence principle should apply for the work distribution thus obtained  \cite{Jarzynski2015}. 
We have proposed a semiclassical expression for the characteristic function that could be applied to general thermodynamic process in chaotic systems.
We have shown analytically that the classical distribution is recovered by taking the effective Planck constant to zero. 
Using numerical simulations for the stadium billiard we show that the semiclassical expression of the work distribution
provides a good approximation to the quantum distribution, and its accuracy is also examined by evaluating the Jarzynski identity. 
We also show that the quantum distribution is best approximated for high temperatures,
and processes that are close to the quench (small values of $\tau$ in our formulation). 
Thus, we also provide further justification for the definition of quantum work by the two-point measurement scheme.

\acknowledgments
The authors have received funding from CONICET (grant nos. PIP 114-20110100048 and 
PIP 11220080100728), ANPCyT (grant nos. PICT-2013-0621 and PICT 2014-3711) and UBACyT. The authors would like to thank M. Saraceno for insight and discussions.
\bibliographystyle{apsrev4-1}
\bibliography{refs-1}
\end{document}